\documentclass[12pt]{article}
\usepackage[centertags]{amsmath}
\usepackage{amsfonts}
\usepackage{amssymb}
\usepackage{amsthm} 
\usepackage{newlfont}
\usepackage{epsfig}
\usepackage{amscd}
\usepackage{graphicx}
\usepackage{epsfig}
\usepackage{amscd}

\newcommand{\beq}{\begin{equation}}
\newcommand{\eeq}{\end{equation}}
\newcommand{\ba}{\begin{array}}
\newcommand{\ea}{\end{array}}
\newcommand{\bea}{\begin{eqnarray}}
\newcommand{\eea}{\end{eqnarray}}
\begin{document}

\begin{center}
{\large \sc \bf { Systems with stationary distribution of quantum correlations:
open 
spin-1/2 chains with $XY$ interaction }
}

\vskip 15pt

{\large 
 E.B.Fel'dman and A.I.~Zenchuk 
}

\vskip 8pt

{\it Institute of Problems of Chemical Physics, RAS,
Chernogolovka, Moscow reg., 142432, Russia},\\
 e-mail:  efeldman@icp.ac.ru, zenchuk@itp.ac.ru 

\end{center}


\begin{abstract}
Although  quantum correlations in a quantum system are characterized by the
evolving quantities (which are  entanglement and discord usually), 
we reveal such basis (i.e. the set of virtual particles) for the
representation of the  density matrix  that the entanglement and/or  discord
between any two virtual particles in such representation are stationary.
In particular, dealing with the nearest neighbor approximation, this system of
virtual particles is represented by the $\beta$-fermions  of the
Jordan-Wigner transformation.  
Such systems are important in quantum information devices because the evolution
of quantum entanglement/discord leads to the  problems of realization of quantum
operations.
The advantage of stationary entanglement/discord is that they are completely
defined by the initial density matrix and by the Hamiltonian governing the
quantum dynamics in the system under consideration.
Moreover, using the special initial condition together with the special system's
geometry,
we construct large cluster of virtual particles with the same pairwise
entanglement/discord.  In other words, the measure of quantum correlations is
stationary in this system and correlations are uniformly "distributed" among all
virtual particles. 
As examples, we use both homogeneous and non-homogeneous spin-1/2 open chains
with  XY-interaction although other types of interactions might be also of
interest.
\end{abstract}


\section{Introduction}
\label{Sec:Introduction}
An attractive problem in quantum information processing is that of revealing of
quantum correlations in a system. 
Presently, two measures  are most acknowledged  as  characteristics of quantum
correlations: the quantum entanglement \cite{Wootters,HWootters,P,AFOV,DPF} and 
the quantum discord 
\cite{Z0,HV,OZ,Z}. 
The so-called  quantum entanglement 
must be noted as  the first quantitative measure of  quantum correlations.
However, it was shown that the  quantum discord  
involves more quantum correlations. As justification of this statement, there
are quantum systems  with zero entanglement  revealing the non-zero discord.
Therefore it is considered \cite{DSC} that namely discord is a proper
measure   responsible for advantages of quantum information devices  (quantum
speed-up and others).

In general, the calculation of discord is a very cumbersome optimization
problem. In spite of  intensive study of discord,
only very special cases have been treated analytically  \cite{L,ARA,Xu,BXH}. 
Nevertheless, namely these cases  correspond to the reduced binary density
matrix in spin-1/2 chains governed by  different Hamiltonians with either the
thermal equilibrium initial state \cite{G,KZ}, 
the initial state with the single excited node \cite{Bose}
and the  initial state with the single polarized node \cite{ZME,FBE,FZ_2012}.

Let us notice, that 
the problem of a proper initial state is one of the fundamental problems in
study of  quntum correlations in different physical systems because of the
technical difficulties in realization of  a particular state. The so-called
thermal equilibrium initial state \cite{G}
is most popular because of its  simple realization. However, the initial state
with a single excited node is more relative, for instance, to quantum
communication lines \cite{Bose,CDEL,ACDE,KS,GKMT,FKZ}. Another example is the 
state with a single polarized node which was produced experimentally \cite{ZME}.
The evolution of a quantum system  with this initial state  at high temperatures
was studied, for instance, in \cite{FBE}, where the quantum echo was found.   

In spite of intensive study, the problem of identification of quantum
correlations is not resolved yet. In particular, the measure of quantum
correlations depends on the basis which is taken for the density matrix
representation. The reason is that, considering different bases, we involve 
different types of  virtual particles. A possible way to avoid this
ambiguity is suggested in ref.\cite{Z_a}, where the unitary invariant discord is
introduced. This measure  takes into account correlations among all possible
virtual particles so that there is no privilege of any particular density
matrix representation.  

 In \cite{FZ_2012}, instead of counting the correlations among all virtual
particles (like in the unitary invariant discord \cite{Z_a}),  the problem of
preferable virtual particles was formulated. Namely, the quantum correlations
among three types of particles are considered separately and compared with each
other. These particles  are following: (i) the fermions, which appear in a
spin-1/2 system with the nearest neighbor interaction after the diagonalization
via  the Jordan-Wigner transformation \cite{JW}  (we call them as the
$\beta$-fermions), (ii) the fermions which are the  Fourier representations of
the $\beta$-fermions (the $c$-fermions),  and  (iii) the physical spin-1/2
particles with the basis of  eigenvectors of the operators $I_{jz}$
($z$-projection of the $j$th spin, $j=1,\dots, N$).  It is shown, that the
distributions of quantum correlations among eigenstates corresponding to three
above bases are completely different. Remind that the $\beta$-representation is
most attractive  owing 
to
its  several remarkable  properties. First 
of all, it yields the   stationary pairwise discord (i.e. the discord between
any two nodes $n$ and $m$), which might be convenient for the realization of
quantum operations. Second, the discord  might be  nonzero even between the
states with zero
entanglement, which confirms the privilege of the discord as a measure of
quantum correlations. Third, the stationary discord is completely  defined by
the initial density matrix (for the given type of quantum interactions)
\cite{FZ_2012}, which provides a simple tool to handle the stationary discord
distribution. Thus, if the first node in the odd-node spin chain is initially
polarized, then 
all nodes of the chain are correlated and the pairwise stationary discord
increases to the center node of the chain \cite{FZ_2012}. If the middle node is
initially polarized, then the system of odd nodes forms a cluster of correlated
fermions with equal pairwise discord. This is the remarkable fact which was not
observed  in the systems of real physical  particles and may be useful for
formation of large quantum registers. 

It is interesting to note that
the dependence of quantum correlations on the particular basis of eigenstates is
considered in  \cite{DJ,DJD,AJDD,DAJD,AJDD2,Lych} from another standpoint.
Namely, the whole space of quantum states of a given  system (the open spin-1/2
chain in the above case) may be splitted into two subspaces. The quantum
correlations are considered in the first one (the subsystem $A$) while another
subsystem $B$ is 
 refereed to as  the  environment. In the above references, the dependence of
quantum correlations on the particularly selected subsystem of quantum states is
demonstrated. In our case the subsystem  $A$   is represented by  the
eigenstates of two virtual particles  of a particular density matrix
representation, while the eigenstates of the rest of particles"  form the 
  environment.

 This paper is devoted to the problem of study  of such system of virtual
particles in a given quantum system that  possesses the stationary discord. We
substantially extend a particular rather qualitative result of
ref.\cite{FZ_2012} concerning the stationary discord in the system of
Jordan-Wigner $\beta$-fermions
 corresponding to  the single initially polarized node in a homogeneous spin-1/2
chain governed by the  XY Hamiltonian  with the  nearest neighbor interactions. 
Namely
 \begin{enumerate}
 \item
Along with the single initially polarized node, we consider the single initially
excited node in a spin-1/2 chain. 
 \item
 We show analytically that the pairwise discord/entanglement are stationary in
the system of virtual particles corresponding to the eigenstates of the
Hamiltonian if only the initial state with the single excited/polarized node is
considered. If we deal with the nearest neighbor interactions, then this
virtual particles are the  $\beta$-fermions of the Jordan-Wigner
representation, which agrees with ref.\cite{FZ_2012}.
 \item
 We find out that 
 both entanglement and  discord are stationary and  nonzero in the above  basis 
 if  the initial state with the single  excited node is taken  
 (remember that the entanglement is zero in the $\beta$-fermion system
considered in ref.\cite{FZ_2012}).
 \item
 We represent  the detailed analytical and numerical  study of the
discord/entanglement distribution 
 in  dependence on the position of the initially excited/polarized node in the
chain. Subsystems  with (almost) uniform pairwise discord/entanglement
distribution have been revealed with analytical formulas for some of them.
Examples of large subsystems are among them.
 \item
 We refer to the inhomogeneous chains  (alternating, 3-alternating and
completely inhomogeneous chain of ref.\cite{CDEL}) and have found several
peculiar subsystems with nonzero discord/entanglement. The diamerization effect
is studied in the alternating chain.
 \item
 Along with the approximation of nearest neighbor interaction, we consider the
case of dipole-dipole interactions among  all nodes (the case of a single
initially excited node) and show that the remote interactions do not
significantly deform the overall pairwise discord/entanglement distribution.
Emphasize that this is an important advantage 
 in comparison with the discord/entanglement in the system of usual spin-1/2
particles, where the remote interactions crucially change this distribution. The
reason is that the remote interactions significantly affect  the spin dynamics
and, consequently, on the dynamics of quantum correlations. However, these
correlations are stationary in our system of virtual particles. 
 \end{enumerate}

 Systems with the stationary discord/entanglement are  important for
construction of the  quantum information devices where the stationary
distribution of quantum correlations simplifies the realization of quantum
operations.
 The matter is that the quantum operations in a given cluster of  correlated
particles may be performed only during the period of  its
 existence (which is defined by the decoherence time associated with a given 
quantum system) and only provided that the quantum correlations are properly
distributed  among all nodes of a cluster. However, 
 even if the quantum correlations are properly distributed at some instant 
 $t_0$, this distribution will be destroyed owing to the quantum evolution. 
Alternately, in a system with the stationary discord,
 we only have to take care about the  proper initial quantum correlations.
Consequently, we receive the relatively simple tool to handle the quantum
correlations varying the initial state and perhaps the type of quantum
interactions in a system.  
 
 Note that the nodes in the  system of the above virtual particles with
stationary pairwise  discord/entanglement are not localized in the physical
space, which makes obstacles in organization of the   impact  on  the state of a
 particular virtual particle using the classical environment. For this reason,
the interface between the operator and quantum device  must be significantly
modified, which is a subject of further study. 
 However, all representations are equivalent from the standpoint of interactions
inside of a quantum system. Thus, we assume that the systems of virtual
particles with stationary distribution of discord/entanglement will be useful
in organization of  those parts of quantum algorithms where the interaction with
the operator is absent ("inner" quantum algorithms).
 
 The paper is organized as follows. 
 In Sec.\ref{Section:gen}, we formulate  general statements on the existence of
systems of virtual particles with the stationary pairwise  discord  in an
arbitrary quantum system. 
Generalizing the idea of ref.\cite{FZ_2012}, we show that the stationary
entanglement/discord  is associated with the system of virtual particles whose
eigenstates diagonalize the Hamiltonian governing the dynamics of a quantum
system. 
 In Sec.\ref{Section:dyn}, we consider the 
spin dynamics in the spin-1/2 system governed by the XY-Hamiltonian and reveal 
general properties of the stationary entanglement/discord in this case.
Then, in Sec.\ref{Section:num}, using the  numerical simulations, we  construct
the stationary pairwise discord distributions  among the virtual particles in
the open spin-1/2 chain  of $N=41$ nodes  (odd $N$) governed by the XY
Hamiltonian
using two types of initial conditions: (i) a single initially excited node and
(ii) a single initially polarized node. 
 In the  case of a single initially excited node, we consider both the
approximation of nearest neighbor interactions and the model with the
dipole-dipole interactions (DDIs)  among all nodes and demonstrate that the
later does not significantly  deform the distribution of the stationary pairwise
quantum entanglement/discord in the system. In the  case of a single initially
polarized node, we consider only the nearest neighbor approximation. In this
case the entanglement is zero for the long chains $N>4$ \cite{FZ_2012} so that
the discord is a proper measure of quantum correlations in this case. We discuss
our results in Sec.\ref{Section:conclusions}. Some auxiliary calculations are
given in the Appendix, Sec.\ref{Section:app}.

\section{Basis of virtual particles with stationary pairwise discord}
\label{Section:gen}

The discord and entanglement in a quantum system are evolving quantities in
general. Their evolution is determined by the Hamiltonian $H$ governing the
dynamics of a quantum system. 
However, there is a basis of virtual particles possessing the stationary 
discord.
Below we consider the Hamiltonian commuting with the $z$-projection of the total
spin 
momentum $I_z$ and show that 
 such basis is that of eigenvectors of Hamiltonian $H$ provided that one of two
following types of initial density matrices $\rho_0$ is considered: (i) the
initial state with a single excited spin and (ii) the initial state with a
single polarized spin.

First, we represent the evolution of the density matrix as
\begin{eqnarray}\label{rhoevH}
\rho(t)=e^{-i Ht} \rho_0 e^{iHt},
\end{eqnarray}
where $\rho_0$ is the initial density matrix.
Diagonalizing $H$ we have
\begin{eqnarray}
H=U \Lambda U^+,
\end{eqnarray}
where $\Lambda$ is the  diagonal matrix of eigenvalues of the Hamiltonian $H$
and $U^+$ is the matrix of its  eigenvectors. In the basis of these 
eigenvectors,
the evolution of the density matrix reads
\begin{eqnarray}
&&\rho^H(t) = \hat E \rho^H_0 \hat E^+, \;\; \rho^H_0  =U^+ \rho_0 U,
\;\; \hat E =e^{-i \Lambda t}.
\end{eqnarray}
To proceed further one has to  fix a particular initial density matrix $\rho_0$.

 \subsection{Single initially excited node in system of spin-1/2 particles}
 \label{Section:one_exc}
 In this section we derive the formulas for the stationary entanglement/discord 
 in a system of spin-1/2 particles  with a single initially excited spin. The
dynamics of the quantum system of $N$ nodes governed by any Hamiltonian
commuting with $I_z$ (the  $z$-projection of the total spin)  can be described
in the $N$-dimensional basis $|n\rangle$, $n=1,\dots,N$, where $n$ means that
$n$th spin is excited (i.e. directed opposite to the strong magnetic field)
while other spins are arranged along the magnetic field. 
The initial density matrix $\rho_0$ corresponding to the   $j$th  initially
excited spin is defined by its elements  as
 \begin{eqnarray}
(\rho_0)_{nm}=\delta_{nj}\delta_{mj}.
 \end{eqnarray}
 Then we can write 
 \begin{eqnarray}
 (\rho^H_0)_{nm} = U_{jn}^* U_{jm}.
 \end{eqnarray}
 As a consequence, we have the relation
 \begin{eqnarray}\label{rho0}
| (\rho^{H}_0)_{nm}|^2 =  (\rho^{H}_0)_{nn}(\rho^{H}_0)_{mm}  .
 \end{eqnarray}
 Since the evolution of the density matrix elements $\rho^{H}_{nm}$ reads as  
 \begin{eqnarray}
 \rho^{H}_{nm}(t) = (\rho^{H}_0)_{nm} \exp(-i (\Lambda_n -\Lambda_m) t),
 \end{eqnarray}
then, taking into account  eq.(\ref{rho0}), we have 
 \begin{eqnarray}\label{rHH}
 | \rho^{H}_{nm}|^2 =  \rho^{H}_{nn}  \rho^{H}_{mm} ,
 \end{eqnarray}
 so that the diagonal elements do not evolve as well as $|\rho^{H}_{nm}|$ for
any $n$ and $m$. This property of the density matrix $\rho^{H}$ results  in the
stationary discord and entanglement.
 
Next, we reduce the density matrix $\rho^{H}$   with respect to all nodes except
for the $n$th and $m$th ones. Emphasize that now we deal with the system of
virtual particles rather then with the system of spin-1/2 particles. 
Introduce the  standard notations for the basis of two particles
\begin{eqnarray}\label{basis}
\{|00\rangle,|01\rangle,|10\rangle, |11\rangle\}, 
 \end{eqnarray}
 where $n$ and $m$ in $|nm\rangle$ mean the different 
 filling numbers for the fermion-like particles. In this basis, the reduced
density matrix reads \cite{DFZ}:
 \begin{eqnarray}\label{rhonm}
 \rho^{(nm)} = \left(
 \begin{array}{cccc}
\sigma_{nm} &0&0&0\cr
0&\rho_{nn}&\rho_{nm}&0\cr
0&\rho_{mn}&\rho_{mm}&0\cr
0&0&0&0
 \end{array}
 \right),\;\;\sigma_{nm}=\sum_{i\neq n,m} \rho_{ii} = 1-\rho_{nn} -
\rho_{mm},\;\;n\neq m.
 \end{eqnarray}
 Note that the last zero in the main diagonal of the reduced density matrix
$\rho^{(nm)}$ appears because the single node was excited initially and the
total projection $I_z$  commutes with the XY Hamiltonian.
 
\subsubsection{Concurrence}
We characterize the entanglement  by the Wootters criterion in terms of the
concurrence \cite{HWootters,Wootters}.
According to \cite{HWootters,Wootters}, 
one needs to construct the spin-flip density matrix
\begin{eqnarray}\label{trho2}
\tilde\rho_{(nm)}(\tau) =(\sigma_{y}\otimes \sigma_y) (\rho^{(nm)})^*(\tau)
(\sigma_y\otimes \sigma_y),
\end{eqnarray}
where the asterisk denotes the complex conjugation 
and the Pauli matrix
$\sigma_y= 2 I_y$. The concurrence for the density matrix $\rho_{(nm)}(\tau)$ is
equal to 
\begin{eqnarray}\label{C}
C=\max(0,2\lambda
-\lambda_1-\lambda_2-\lambda_3-\lambda_4),\;\;\lambda=\max(\lambda_1,\lambda_2,
\lambda_3,\lambda_4),
\end{eqnarray}
where $\lambda_1$, $\lambda_2$, $\lambda_3$ and $\lambda_4$ are the square roots
of the eigenvalues of the matrix product 
$\rho_{(nm)}(\tau) \tilde \rho_{(nm)}(\tau)$.
For the density matrix $\rho$ given by eq.(\ref{rhonm}) we have only one nonzero
$\lambda$:
\begin{eqnarray}\label{lambdas}
&&
\lambda=\lambda_1=2 \sqrt{\rho_{nn}\rho_{mm}} 
\end{eqnarray}
Substituting eqs.(\ref{lambdas}) into eq.(\ref{C}) we obtain
\begin{eqnarray}\label{C2}
C_{nm}=\max\left(0,2\sqrt{\rho_{nn}\rho_{mm} } \right),\;\;n\neq m.
\end{eqnarray}

\subsubsection{Discord}
The matrix (\ref{rhonm}) is so-called the X-matrix whose discord has been
studied in \cite{ARA}. Although this reference contains a mistake concerning 
the number of 
arbitrary optimization parameters in the calculation of the classical part of mutual 
correlations (see erratum in ref.\cite{ARA} and ref.\cite{H}), this mistake 
has no value in our case because the element
$\rho_{14}$ is zero in  all  density matrices considered below and relation (\ref{rHH}) holds. 
As a consequence, we have only  one optimization
 parameter, which we denote by $\eta$ (see eqs.(\ref{p},\ref{theta})). Thus, 
we use the algorithm 
developed in the above reference for the calculation of discord.
 Remind that  the discord between the  particles $n$ and $m$  of a biparticle 
quantum system 
 may be calculated as 
 \begin{eqnarray}\label{Q}
Q_m={\cal{I}}(\rho) -{\cal{C}}^m (\rho),
\end{eqnarray} 
provided that the von Neumann type measurements are performed  over the particle
$m$. 
Here   ${\cal{I}}(\rho)$ is the total mutual information \cite{OZ} which may be
written as follows: 
\begin{eqnarray}\label{I}
&&
{\cal{I}}(\rho) =S(\rho^{(n)}) + S(\rho^{(m)}) + \sum_{j=0}^1 \lambda_j \log_2
\lambda_j,\\\nonumber
\end{eqnarray}
where  $\lambda_j$ ($j=0,1$) are  the non-zero eigenvalues of the density
matrix $\rho^{(nm)}$,
\begin{eqnarray}
\lambda_0= \rho_{mm}+\rho_{nn},\;\;\lambda_1=1-\lambda_0,
\end{eqnarray}
$\rho^{(n)}={\mbox{Tr}}_m \rho^{(nm)}$ and $\rho^{(m)}={\mbox{Tr}}_n
\rho^{(nm)}$ are the reduced density matrices
and the appropriate entropies $S(\rho^{(n)})$ and $S(\rho^{(m)})$ are given by
the following formulas:
\begin{eqnarray}\label{SAB}
&&S(\rho^{(n)})=-(1- \rho_{mm} ) \log_2(1-\rho_{mm}) -
          \rho_{mm} \log_2\rho_{mm} ,\\\nonumber
&&S(\rho^{(m)})=-(1-\rho_{nn} ) \log_2(1-\rho_{nn}) -
 \rho_{nn} \log_2\rho_{nn} .
\end{eqnarray}
The so-called classical counterpart ${\cal{C}}^B (\rho^{(nm)})$ of the mutual
information 
can be found considering the minimization over the projective measurements performed 
on
the subsystem $B$ as follows \cite{ARA}:
\begin{eqnarray}\label{CB2}
&&
{\cal{C}}^{(m)} (\rho)=S(\rho^{(n)}) -\min\limits_{\eta\in[0,1]}(p_0 S_0 + p_1
S_1),
\end{eqnarray}
where
\begin{eqnarray}\label{S}
&&S(\theta_i)\equiv S_i = -\frac{1-\theta_i}{2}\log_2\frac{1-\theta_i}{2}-
                 \frac{1+\theta_i}{2}\log_2\frac{1+\theta_i}{2},
\\\label{p}
&&p_i=\frac{1}{2} \Big(1+(-1)^i\eta(1-2\rho_{nn}) \Big),\\\label{theta}
&&\theta_i=\frac{1}{p_i}\Big[(1-\eta^2) \rho_{nn}  \rho_{mm} +\\\nonumber
&&
\frac{1}{4}
\Big(
1-2\rho_{mm} +(-1)^i \eta(1-2(\rho_{nn}+\rho_{mm}))\Big)^2 
 \Big]^{1/2},
\\\nonumber
&&
i=0,1.
\end{eqnarray}
Here we introduce the parameter $\eta$ instead of $k$ in \cite{ARA},
$k=(1+\eta)/2$.
It is simple to show that the quantum discord $Q_n$ obtained performing 
 the von Neumann type measurements  on the particle $n$ can be calculated 
as follows:
\begin{eqnarray}\label{QA}
Q_n=Q_m|_{\rho^{(nn)}\leftrightarrow \rho^{(mm)}}
\end{eqnarray}
for the system with the density matrix $\rho^{(nm)}$ given by eq.(\ref{rhonm}).
Then we define the  discord $Q_{nm}$ as the minimum of $Q_{n}$ and $Q_{m}$
\cite{FZ} 
\begin{eqnarray}\label{def_discord}
Q_{nm}=
\min(Q_n,Q_m), \;\; n\neq m
\end{eqnarray}
with the obvious property $Q_{nm}=Q_{mn}$.
We see that, since  $\rho_{nn}$ and  $\rho_{mm}$ do not depend on time, 
the discord does not evolve as well.

Similar to \cite{FZ_2012}, we can  show that the minimum in eq.(\ref{CB2})
corresponds to $\eta=0$ (the proof of this  statement is given in Appendix
\ref{Section:app}), so that 
we result in the following explicit formula for the discord between any two
nodes: 
\begin{eqnarray}\label{final_discord_ex}
&&
Q_m=1 -\rho_{nn} \log_2 \rho_{nn} - (1-\rho_{nn})\log_2(1-\rho_{nn})
+(\rho_{nn}+\rho_{mm}) \log_2 (\rho_{nn}+\rho_{mm})+\\\nonumber
&&
 (1-\rho_{nn}-\rho_{mm})\log_2(1-\rho_{nn}-\rho_{mm})  -\\\nonumber
 &&\frac{1}{2}
 \Big(1-\sqrt{1-4\rho_{mm}(1-\rho_{nn}-\rho_{mm})}\Big)\log_2
(1-\sqrt{1-4\rho_{mm}(1-\rho_{nn}-\rho_{mm})}) -
 \\\nonumber
 &&
 \frac{1}{2}
 \Big(1+\sqrt{1-4\rho_{mm}(1-\rho_{nn}-\rho_{mm})}\Big)\log_2
(1+\sqrt{1-4\rho_{mm}(1-\rho_{nn}-\rho_{mm})})
\end{eqnarray}
We see that  both the discord $Q_m$ and the concurrence $C_{m}$ are zero if
either $\rho_{nn}$ or $\rho_{mm}$ is zero.

  \subsection{Single initially polarized  node}
  \label{Section:pol}
  
The initial state with a single excited spin considered in 
Sec.\ref{Section:one_exc} is hard for the realization in the experiment and is
associated with low temperatures. 
On the contrary,   the initial  state with a  single polarized node is
realizable even at  high temperatures \cite{ZME,FBE}. This is a motivation 
to consider the discord in a chain with the initially polarized spin.

The stationary discord in the homogeneous spin-1/2 chain  with 
single initially polarized node  governed by the XY-Hamiltonian was introduced 
in \cite{FZ_2012}. Here we represent the more detailed analysis 
of that case and 
generalize results on the  non-homogeneous spin-1/2 chain keeping the
approximation of nearest neighbor interactions.

For a non-homogeneous chain,
similar to ref.\cite{FZ_2012},  we  take advantage of the Jordan-Wigner
transformation \cite{JW}. Let us emphasize that this transformation is
applicable to any Hamiltonian 
at the approximation of the nearest neighbor interactions. 
Let
$I_{i\alpha}$ ($i=1,\dots,N$, $\alpha=x,y,z$) be the $i$th spin projection on
the $\alpha$-axis. 
The initial  density matrix $\rho_0$ corresponding to the initial state of the
spin system with the single polarized $j$th node ($1\le j \le N$ ) at arbitrary
temperature reads \cite{FBE,FZ_2012}
\begin{eqnarray}
\rho_0=\frac{e^{\beta I_{jz}}}{Z}=
\frac{1}{2^N}\left(1+2 I_{jz}\tanh \frac{\beta}{2}\right),\;\;Z={\mbox{Tr}}
(e^{\beta I_{jz}}) = 2^N \cosh\frac{\beta}{2},
\end{eqnarray}
where $\beta=\frac{\hbar \omega_0}{kT}$ is the dimensionless inverse
temperature, $\hbar$ is the Planck constant, $k$ is the Boltzmann constant, and
$T$ is the temperature of the system. 
The evolution of the density matrix 
reads:
\begin{eqnarray}\label{rho_t}
\rho(t)= e^{-i t H} \rho_0 e^{i t H}= 
\frac{1}{2^N} e^{-iH t} (1+2 I_{jz} \tanh\frac{\beta}{2}) e^{iH t}.
\end{eqnarray}
Diagonalizing the 
 Hamiltonian 
 using  
 the Jordan-Wigner transformation method \cite{JW} we result in the following
operator representation of the Hamiltonian $H$:
\begin{eqnarray}
H=\sum_{k} \varepsilon_k \beta_k^+\beta_k,
\end{eqnarray}
where the fermion operators $\beta_k$ are defined in terms of  the other fermion
operators $c_j$  by means of the  transformation (which reduces to  the Fourier
transformation in the case of homogeneous spin-1/2 chain)
\begin{eqnarray}
\beta_k = \sum_{j=1}^N U_{kj} c_j,
\end{eqnarray}
and the fermion operators  $c_j$ are defined as \cite{JW}
\begin{eqnarray}
c_j=(-2)^{j-1} I_{1z}I_{2z}\dots I_{z(j-1)} I^-_j.
\end{eqnarray}
Here the eigenvalues $\varepsilon_k$ and the matrix of eigenvectors  $U_{kj}$
depend on the particular Hamiltonian. 
Then the density matrix (\ref{rho_t})
can  be transformed to the following form \cite{FBE}
\begin{eqnarray}\label{rhot}
\rho(t)=\frac{1-\tanh\frac{\beta}{2}}{2^N} +
\frac{\tanh\frac{\beta}{2}}{2^{N-1}} \sum_{k,k'}
e^{-i t(\varepsilon_k-\varepsilon_{k'})} U_{kj} U_{k'j} \beta^+_k\beta_{k'}
\end{eqnarray}
(here we take into account  reality of $U_{kj}$).
Similar to \cite{FZ_2012}, we will study the quantum correlations between any
two $\beta$-fermions. 

The first step in calculation of the discord  between the  $n$th and $m$th nodes
 is the construction of the 
 reduced density matrix with respect to all nodes except for the $n$th and $m$th
ones.
 We use notations (\ref{basis})
for the  vectors of the $\beta$-fermion basis.
Reducing  density matrix (\ref{rhot}) we obtain:
\begin{eqnarray}
\rho^{(nm)} = \frac{1}{4} - \frac{\tanh\frac{\beta}{2} }{4} (U_{nj}^2 +U_{mj}^2)
+
\frac{\tanh\frac{\beta}{2} }{2} \sum_{k,k'=n,m} e^{-i
t(\varepsilon_k-\varepsilon_k')} U_{kj}U_{k'j} \beta^+_k \beta_{k'}.
 \end{eqnarray}
Its matrix representation in the  basis (\ref{basis}) reads
\begin{eqnarray}
\label{rhonm_matr}
\rho^{(nm)}= \left(
\begin{array}{cccc}
J^\beta_{00}  + J^\beta_{mm}+ J^\beta_{nn}&0&0&0\cr
0&J^\beta_{00}  + J^\beta_{mm}&J^\beta_{mn}&0\cr
0&J^\beta_{nm} &J^\beta_{00}  + J^\beta_{nn}&0\cr
0&0&0&J^\beta_{00}
\end{array}
\right),
\end{eqnarray}
where 
\begin{eqnarray}\label{J_bet}
&&
J^\beta_{00}= \frac{1}{4} - \frac{\tanh\frac{\beta}{2} }{4} (U_{nj}^2 +U_{mj}^2)
,
\\\nonumber
&&
J^\beta_{nm}=\frac{\tanh\frac{\beta}{2} }{2}  e^{-i
t(\varepsilon_n-\varepsilon_m)} U_{nj} U_{mj}
\end{eqnarray}
It is obvious that
\begin{eqnarray}\label{J_nn_beta}
J^\beta_{nn}=\frac{\tanh\frac{\beta}{2} }{2}  U_{nj}^2,
\end{eqnarray}
which does not depend on the time $t$.
Then \cite{FZ_2012}
\begin{eqnarray}\label{finaldiscord}
Q_m&=&-\frac{1}{2}\Big(
(1-2 J_{nn}) \log_2 (1-2 J_{nn})+(1+2 J_{nn}) \log_2 (1+2 J_{nn})-\\\nonumber
&&
(1-2 J_{mm} - 2 J_{nn}) \log_2(1-2 J_{mm} - 2 J_{nn})-\\\nonumber
&&
(1+2 J_{mm} + 2 J_{nn}) \log_2(1+2 J_{mm} + 2 J_{nn})+\\\nonumber
&&
(1-2 \sqrt{J_{mm}(J_{mm}+J_{nn})})\log_2(1-2 \sqrt{J_{mm}(J_{mm}+J_{nn})})
+\\\nonumber
&&
(1+2 \sqrt{J_{mm}(J_{mm}+J_{nn})})\log_2(1+2 \sqrt{J_{mm}(J_{mm}+J_{nn})})\Big).
\end{eqnarray}
The  discord $Q_{nm}$ is defined by  eq. (\ref{def_discord}) with $Q_n$ from
eq.(\ref{QA}).
Similar to  Sec.\ref{Section:one_exc}, the discord $Q_{nm}$ is zero if either
 $J_{nn}$ or $J_{mm}$ is zero. Emphasize, that the formula (\ref{finaldiscord})
for the pairwise discord $Q_{n}$ in the arbitrary non-homogeneous chain
coincides with that derived for the homogeneous chain in \cite{FZ_2012} up to
the definition of elements $J_{nn}$ (\ref{J_nn_beta})
where $U_{nj}$ are the elements of the $j$th eigenvector of the Hamiltonian $H$ 
governing the dynamics of the non-homogeneous chain. 

\section{Dynamics in spin-1/2 chain with XY Hamiltonian and general properties
of stationary discord}
\label{Section:dyn}
We consider a one-dimensional open chain of spin-1/2 particles in the strong
external magnetic field governed by the XY Hamiltonian

\begin{eqnarray}
\label{HamiltonianXY}
&&
{\cal{H}}=\sum_{{i,j=1}\atop{j>i}}^{N}
D_{i,j}(I_{i,x}I_{j,x} + I_{i,y}I_{j,y}),
\;\;\;D_{i,j}=\frac{\gamma^2 \hbar}{2r_{i,j}^3}.
\end{eqnarray}
Here $D_{i,j}$ is the coupling constant between the $i$th and the $j$th nodes.
Hereafter we will  use the dimensionless  time $\tau$, distances $\xi_{n,m}$,
 coupling constants $d_{n,m}$  defined as follows:
\begin{eqnarray}\label{tau}
&&
\tau= D_{1,2} t,\;\;\xi_{n,m}=\frac{r_{n,m}}{r_{1,2}},
\;\;
d_{n,m}=\frac{D_{n,m}}{D_{1,2}}=\frac{1}{\xi_{n,m}^3},\;\;d_{1,2}=1.
\end{eqnarray}
Using  definitions (\ref{tau}), 
the Hamiltonian  (\ref{HamiltonianXY}) may be written as follows:
\begin{eqnarray}
\label{HamiltonianXYdimles}
&&
{\cal{H}}=D_{1,2} H,\;\;\;
H=\sum_{{i,j=1}\atop{j>i}}^{N}
d_{i,j}(I_{i,x}I_{j,x} + I_{i,y}I_{j,y}).
\end{eqnarray}
For the nearest neighbor interaction approximation  we write 
$d_{i,j}= d_i \delta_{j,i+1}$, $j>i$. 

\subsection{Homogeneous chain with nearest neighbor interaction approximation}
\label{Section:1excited}
First, we consider the homogeneous spin-1/2 chain, i.e. $d_i=d\equiv 1$,
$i=1,\dots,N-1$.
 In this case  we have the following  formulas 
for the eigenvalues and eigenvectors of the Hamiltonian \cite{FBE}:
\begin{eqnarray}\label{hom}
\varepsilon_k = \cos\;\frac{\pi k}{N+1},\;\;
U_{kj} = \sqrt{\frac{2}{N+1}} \sin \frac{\pi k j}{N+1},
\end{eqnarray}
which hold for the  initial state with  both a single excited node and a single
polarized node.
Consequently,
\begin{eqnarray}\label{rho}
\rho_{nn} = U_{nj}^2 = \frac{2}{N+1}  \sin^2 \frac{\pi n j}{N+1},\\\label{J}
J_{nn} = \frac{\tanh\frac{\beta}{2}}{2} U_{nj}^2 = 
\frac{\tanh\frac{\beta}{2}}{N+1}  \sin^2 \frac{\pi n j}{N+1},
\end{eqnarray}
These expressions must be substituted  into eqs.(\ref{C2}),
(\ref{final_discord_ex}) and (\ref{finaldiscord}).  The most simple is the
expression for the concurrence  in the case of the XY Hamiltonian with a single
excited node (remember that the concurrence in the case of a single polarized
node is zero for long chains \cite{FZ_2012}):
\begin{eqnarray}\label{CJ}
C_{nm}(j) = \frac{4}{N+1}\left|\sin \frac{\pi n j}{N+1}\sin \frac{\pi m
j}{N+1}\right|.
\end{eqnarray}
Now we reveal some properties of the pairwise entanglement/discord distribution
among the virtual particles.
We can always write
\begin{eqnarray}
\frac{j}{N+1}=\frac{m_1}{m_2} ,\;\;
\end{eqnarray}
where $m_1$ and $m_2$ are integers.
If $m_2<N+1$, i.e. the integers $j$ and $N+1$ have the common factor, 
then the discord and the concurrence reveal the periodic behavior with zeros at
such nodes $n$  that
\begin{eqnarray}
\frac{n j}{N+1} =1.
\end{eqnarray}
In  the
periodic case, the concurrence $C_{nm}$ and/or the discord  $Q_{nm}$   take
several different  values depending  on $n$ and $m$. The number of these values
is defined by the number of different pairs of values of $\rho_{nn}$ and
$\rho_{mm}$ (\ref{rho}), $n,m=1,\dots,N$. 

For our convenience, let us use superscripts $ex$ and $pol$ to mark quantities
associated with the initially excited and the initially polarized single node 
respectively,
i.e. we will write $Q^{ex}_{nm}$,
$Q^{pol}_{nm}$, $C^{ex}_{nm}$ (while $C^{pol}_{nm} \equiv 0$ for long chains).
Next, we  formulate several statements on existence of large clusters of nodes
with equal pairwise discord (concurrence) for odd $N$.
\begin{enumerate}
\item
If $N$ is odd and $j=(N+1)/2$, then eq.(\ref{rho}) yields only one   non-zero
value for $\rho_{nn}$:
\begin{eqnarray}\label{rho_1}
\rho_1\equiv \rho_{nn}=
\frac{2}{N+1},\;\;\; n=1,3,5,\dots.
\end{eqnarray}
In this case the  nonzero pairwise  discord appears  only among the odd nodes
and its value is the same for any pair of odd nodes:
\begin{eqnarray}\label{Q_1}
Q^{ex,pol}_1\equiv
Q^{ex,pol}_{2k_1+1,2k_2+1}=Q^{ex,pol}_{13},\;\;k_1,k_2=1,3,\dots,\;k_2>k_1
\end{eqnarray}
 (we do not represent the explicit formula for discord).
The concurrence reads in this case:
\begin{eqnarray}\label{C_1}
C^{ex}_1\equiv C^{ex}_{2k_1+1,2k_2+1}=C^{ex}_{13}
=\frac{4}{N+1},\;\;k_1,k_2=0,1,2,\dots,\;k_2>k_1.
\end{eqnarray}
\item
If $N=5 + 6i$, $i=1,2,\dots$,
and $j=\frac{N+1}{3}=2(i+1)$, then again we have only one nonzero value for
$\rho_{nn}$,
\begin{eqnarray}\label{rho_2}
\rho_1\equiv \rho_{nn}=
\frac{2}{N+1}\sin^2\frac{\pi}{3}= \frac{3}{2(N+1)},& n = 3k-1,
\;3k-2,\;\;k=0,1,2,\dots,2 i +2,
\end{eqnarray}
The nonzero discord will be only among the nodes from the set $\{3k-1,
\;3k-2,\;k=1,\dots, 2 i +2\}$ and is the same for any pair from this set. It
reads:
\begin{eqnarray}\label{Q_2}
Q^{ex,pol}_1=Q^{ex,pol}_{12},
\end{eqnarray}
and is given by
eqs.(\ref{QA},\ref{def_discord},\ref{final_discord_ex},\ref{finaldiscord}) with
proper substitutions for $J_{ii}$, $\rho_{ii}$, $i=n,m$.
The concurrence between any pair from this set reads:
\begin{eqnarray}\label{C_2}
C^{ex}_1=C^{ex}_{12}=\frac{3}{N+1}.
\end{eqnarray}
\item
In general, if
$\frac{j}{N+1}=\frac{m_1}{m_2}$, then 
\begin{eqnarray}\label{rho_3}
\rho_{nn}=\frac{2}{N+1}\sin^2 \frac{\pi n m_1}{m_2},
\end{eqnarray}
which is nonzero if the ratio $\frac{nm_1}{m_2}$ is not integer. Therewith for
any $n_1$ and $n_2$ such that 
$\frac{n_1 m_1}{m_2} = 1\pm \frac{n_2 m_1}{m_2}$ we have 
$\rho_{n_1n_1}=\rho_{n_2n_2}$.
\end{enumerate}

 \subsection{Alternating chain with odd $N$ and nearest neighbor interaction
approximation}
 \label{Section:alt_ch}
 In this case $d_1=d_{2n-1}=1$, $d_2=d_{2n}$, $n=1,2,\dots$ and we use the
parameter $\delta=d_2/d_1\equiv d_2$ as the dimerization degree.
 It is known that the Hamiltonian is analytically diagonalizable in this case
for both odd \cite{FR} and even $N$ \cite{KF}.
 
 It can be readily shown that,  if $N$ is odd and the excited node $j$ is even,
then the discord coincides with that calculated for  the homogeneous chain
\cite{FZ_2012}. In fact,
using  formulas for the eigenvalues and the eigenvectors of the XY Hamiltonian
derived  in ref.\cite{FR}, we conclude  that both eigenvalues and eigenvectors
of the XY Hamiltonian involved in the calculation of the concurrence/discord in
this case coincide with those used for the calculation of the pairwise discord in
the homogeneous chain, see eqs.(\ref{hom}). If $j$ is odd, then the discord
depends on the dimerization degree  $\delta$. However, we have found that if
$\delta \to 0$ (the limit of the non-interacting  dimers) then the following
expressions for the eigenvalues 
 follow from the formulas of ref.\cite{FR}
 \begin{eqnarray}\label{del_e}
2 \varepsilon_k = \lambda_k=\left\{\begin{array}{ll}\displaystyle 
 d_1, &\displaystyle k=1,2,\dots,\frac{N-1}{2}\cr\displaystyle 
 0,& \displaystyle k=\frac{N+1}{2}\cr\displaystyle 
 -d_1, &\displaystyle k=\frac{N+1}{2} +1,\frac{N+1}{2} +2,\dots, N\cr
 \end{array}\right..
 \end{eqnarray}
 For the elements of the eigenvectors at $k\neq \frac{N+1}{2}$ we have
 \begin{eqnarray}\label{del_U}
U_{kj}= \left\{
\begin{array}{ll}\displaystyle 
 \frac{ d_1}{\lambda_k}\sqrt{\frac{2}{N+1}}
\sin\Big(\frac{
\pi k(j+1)}{N+1} 
\Big)
 ,&j=1,3,5,\dots,N\cr\displaystyle 
\sqrt{\frac{2}{N+1} }\sin\Big(
\frac{\pi k j}{N+1}
\Big),& j=2,4,\dots,N-1
\end{array}\right. ,
\end{eqnarray}
Finally,  at $k= \frac{N+1}{2}$, we obtain
 \begin{eqnarray}\label{del_U2}
U_{j,(N+1)/2} =\left\{
\begin{array}{ll}
1,& j=N\cr
0,& j\neq N
\end{array}\right. .
\end{eqnarray}
Formulas (\ref{del_U},\ref{del_U2}) demonstrate that, in this case,
the eigenvectors corresponding to the odd and even initially excited node $j$
are very similar up to the shift $j\to j+1$. Consequently, for small
dimerization parameter $\delta$, discords corresponding to $j=2 n$ and $j=2n-1$
are very similar.
In addition, the discord vanishes if $j=N$. 

We shall emphasize that,  considering the stationary discord  in the basis of 
the Hamiltonian eigenstates, we obtain quantum correlations even in a system
of non-interacting dimers, $\delta\to 0$. This is consequence of the 
diagonalization process, which involves  all nodes regardless of the values of
their  interactions.
 
 \section{Numerical simulations of the dynamics in  spin chain with odd $N$,
$N=41$}
 \label{Section:num}
 All calculations  of this section are performed for the spin chain with odd
$N$.
 The stationary pairwise discord distribution for the even $N$ is essentially
the same.
Both models  with the nearest neighbor interaction approximation and  with the
dipole-dipole  interactions (DDIs) among all nodes are considered. In this
regard, it is important to note that  the interactions among the remote nodes
effect significantly 
on the entanglement between the spin-1/2 particles and   on the  state transfer
process along the spin-1/2 chains  \cite{FKZ}. However, the stationary discord
is much less sensitive to the remote node  interactions, which only deform the
distribution of the pairwise stationary discord among the nodes, see
Sec.\ref{Section:excited_num}.  This happens because the  time evolution of the
entanglement (and the polarization)
is  very sensitive to the interactions of the remote nodes.  In general, these
interactions speed up the signal propagation. However, the evolution is
"averaged" in the stationary discord so that   
the effect of remote nodes is 
suppressed. 

Hereafter we study the pairwise discord using different initial states with
either single excited or single polarized spin. We say that two nodes are
correlated if the corresponding  concurrence and/or discord are non-vanishing. 

 \subsection{Single excited node: homogeneous spin chain}
 \label{Section:excited_num}

\begin{figure*}
   \epsfig{file=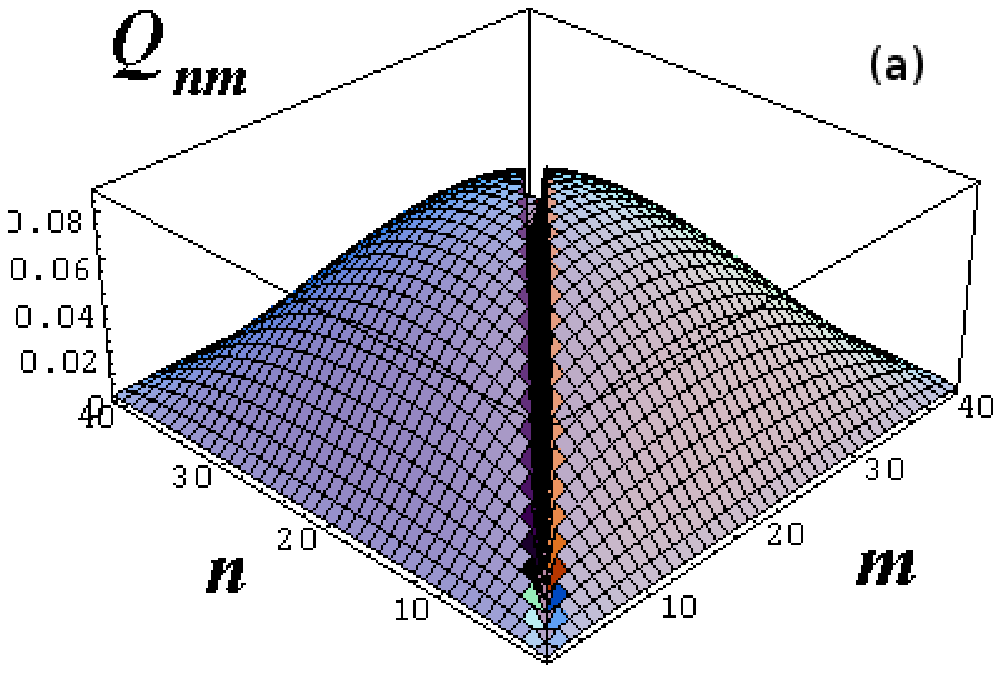,
  scale=0.8
   ,angle=0
}
\epsfig{file=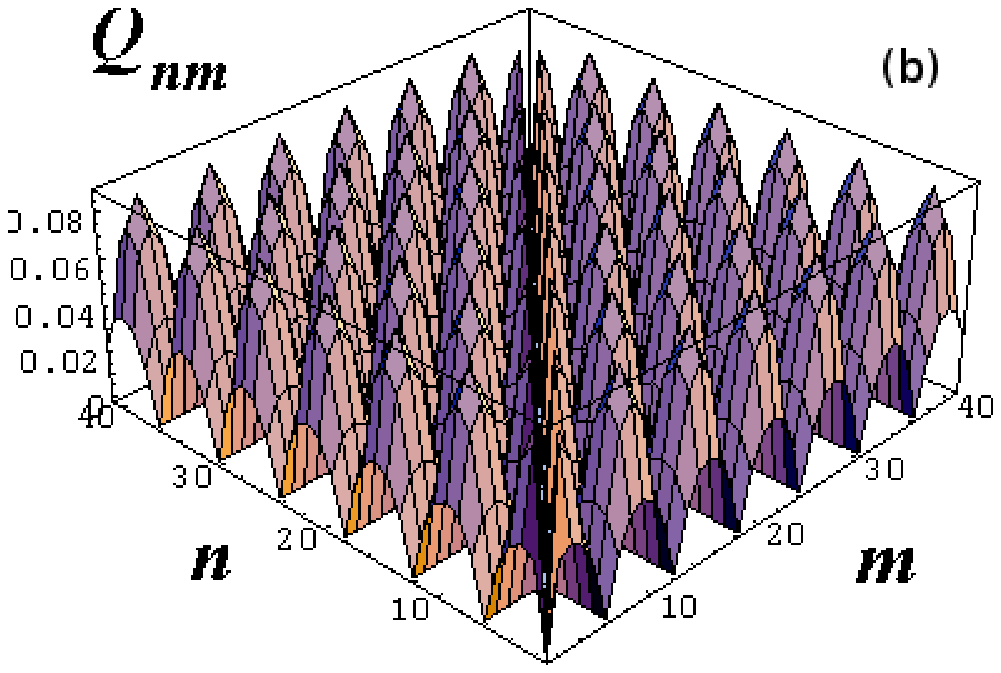, 
    scale=0.8
   ,angle=0
}
  \caption{ The distribution of the stationary  pairwise discord $Q_{nm}$ among
the virtual particles in   the homogeneous spin-1/2 chain with $N=41$ at the
nearest neighbor interaction approximation. For convenience, we take  $Q_{nn}$
equal to zero which is indicated in all pictures below. The initially excited
nodes are following  (a) $j=1$, the discord distribution is bell-shaped (b)
$j=7$, the discord takes 6 different values given in eqs.(\ref{Qex}), nodes
$n=6i$, $i=1,\dots,6$, do not correlate with others} 
  \label{Fig:disc_hom} 
\end{figure*}

Now we apply  formulas (\ref{rho_1}-\ref{rho_3}) to the homogeneous spin chain
of
$N=41$ nodes.
 Because of the symmetry, it is enough to consider the initially excited nodes
$j\le \frac{N+1}{2}$ (for odd $N$).  First we calculate the discord using the
nearest neighbor interaction approximation.  The basic results are following:
 
1. If $j=1$, then all nodes are correlated. Both the discord and  the concurrence 
increase   to the center of the chain of virtual particles, Fig.
\ref{Fig:disc_hom}a. 

2. From eqs.(\ref{rho}-\ref{CJ}) it follows that the discord and the concurrence are
the periodic function of $n$ if $j=6$, $7$, 12, 14, 18, 21. For instance, if 
 $j=7$
(see Fig.\ref{Fig:disc_hom}b)  than we have three different values for 
$\rho_{nn}$:
\begin{eqnarray}\label{rhoexj7}
&&
\rho_{nn}=\rho_1=\frac{1}{84},\;\;n = 6 i +1, 6 i +5,\;\;
i=0,1,\dots,6,\\\nonumber
&&
\rho_{nn}=\rho_2=\frac{1}{28},\;\;n=  6 i +2, 6 i +4,\;\;
i=0,1,\dots,6,\\\nonumber
&&
\rho_{nn}=\rho_3=\frac{1}{21},\;\;n= 6 i +3, \;\;i=0,1,\dots,6.
\end{eqnarray}
They produce 6 different values of the discord
\begin{eqnarray}\label{Qex}
&&
Q^{ex}_1=Q^{ex}_{15}\approx 0.023,\;\;Q^{ex}_2=Q^{ex}_{24}\approx 0.067,\;\;
Q^{ex}_3=Q^{ex}_{39}\approx 0.088,\\\nonumber
&&
Q^{ex}_4=Q^{ex}_{12}\approx 0.036,\;\;Q^{ex}_5=Q^{ex}_{13}\approx 0.040,\;\;
Q^{ex}_6=Q^{ex}_{23}\approx 0.076
\end{eqnarray}
and of the concurrence
\begin{eqnarray}\label{Cex}
&&
C^{ex}_1=C^{ex}_{15}=\frac{1}{42},\;\;C^{ex}_2=C^{ex}_{24}=\frac{1}{14},\;\;
C^{ex}_3=C^{ex}_{39}=\frac{2}{21},\\\nonumber
&&
C^{ex}_4=C^{ex}_{12}=\frac{1}{14\sqrt{3}},\;\;C^{ex}_5=C^{ex}_{13}=\frac{1}{21},
\;\;
C^{ex}_6=C^{ex}_{23}=\frac{1}{7\sqrt{3}}.
\end{eqnarray}
We see that the correlations are most strong among the nodes from the set 
$\{6 i +3; \;\;i=0,1,\dots,6\}$. The set of nodes $\{6 i +2, 6 i +4;\;\;
i=0,1,\dots,6\}$
 is less correlated with the first one. The   correlations  with the set $\{6 i
+1, 6 i +5;\;\; i=0,1,\dots,6\}$ are minimal. 
Nevertheless,
all nodes are correlated accept for the nodes $n=6i$, $i=1,\dots,6$ because all
the pairwise discords involving these nodes  are zeros.

3. If $j=14$, then  we have one nonzero value for the elements  $\rho_{nn}$ with
$n=3i-1, 3i-2$, $i=1,\dots,13$:
\begin{eqnarray}\label{rhoexj14}
\rho_{nn}=\rho_1=\frac{1}{28}
\end{eqnarray}
The appropriate nonzero discord  and concurrence are following:  
\begin{eqnarray}
Q^{ex}_1=0.067, \;\; 
C^{ex}(\rho_1,\rho_1)=\frac{1}{14}.
\end{eqnarray}
Thus, the pairwise concurrences  and/or discords are nonzero  and  take the same
values for any pair of nodes  from the cluster 
$\{3 i-1,3i-2; \; i=1,2,\dots,13\}$.

4. If $j=21$, then, again, there is only one nonzero value  $\rho_{nn}$ for all
odd $n$:
\begin{eqnarray}\label{rhoexj21}
\rho_1\equiv \rho_{nn}=\frac{1}{21} ,\;\;\;
n=2i -1,\;\;i=0,1,\dots,21 .
\end{eqnarray}
 The appropriate values of the discord and concurrence are   
 \begin{eqnarray}
 Q^{ex}_1=0.088,\;\; 
C^{ex}_1=\frac{2}{21}.
\end{eqnarray}
Thus, the pairwise concurrences/discords are nonzero and equal each other  for
any pair   from the family  of odd nodes.

\begin{figure*}

\epsfig{file=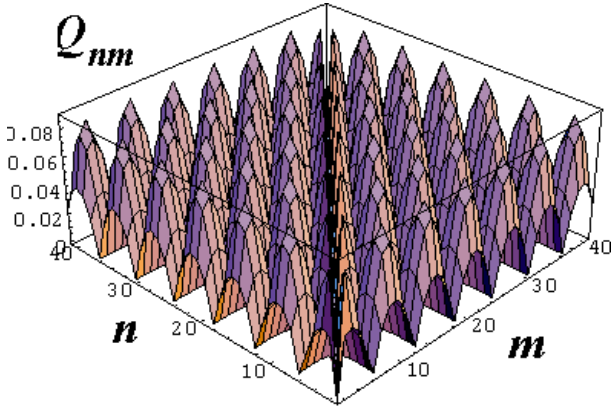,
    scale=0.8
   ,angle=0
}

  \caption{ The distribution of the stationary  pairwise discord $Q_{nm}$ among
the virtual particles in   the homogeneous spin-1/2 chain with $N=41$;
interactions among all nodes are taken into  
 account; the initially excited node is $j=7$. This distribution only slightly
differs from that shown in 
 Fig.\ref{Fig:disc_hom}b}
  \label{Fig:disc_hom_all} 
\end{figure*}

It is remarkable, that the concurrence  has the same distribution among nodes as
discord. For this reason we do not represent the figures with the concurrence
distribution.   In addition, we verify that, involving the interactions among
all nodes, the discord distribution does not become significantly deformed,  which
confirms the arguments  given in the beginning of Sec.\ref{Section:num}.  As an
example, in  Fig.\ref{Fig:disc_hom_all}, we   
represent the discord  distribution corresponding to  the $7$th initially
excited spin (i.e. $j=7$) and the Hamiltonian involving the   DDIs among all
nodes of the spin chain.

\subsection{Single excited node: non-homogeneous spin chains}

In this section we show that varying either the coupling constants in the
Hamiltonian or the initially excited node we may handle the size of the cluster
of the correlated particles.
Having this possibility, we may select the cluster of required nodes from the
whole  chain of virtual particles which is necessary for flexibility of the
quantum algorithms. 
In this regards we notice that the problem of variation of  the coupling
constants may be effectively resolved using, for instance, the optical lattice 
\cite{PK}. In addition, the effect of variable coupling constants in  spin
chains may be effectively replaced with the  variable
 magnetic field \cite{DZ} surrounding the spin chain.

\subsubsection{Alternating spin chain}
\label{Section:alt}
\begin{figure*}
\epsfig{file=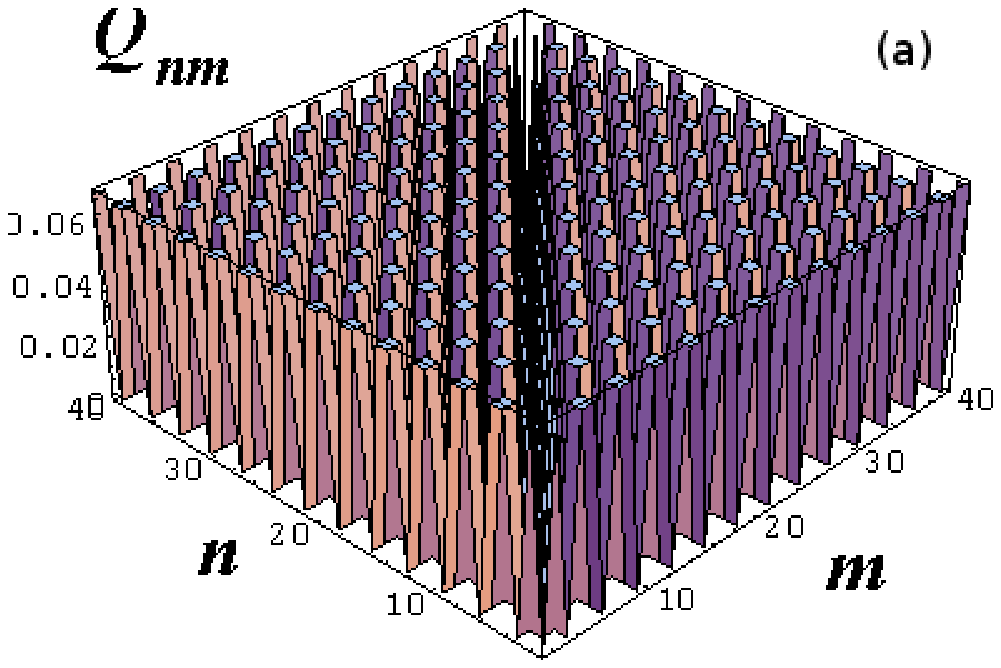,
    scale=0.8
   ,angle=0
}
\epsfig{file=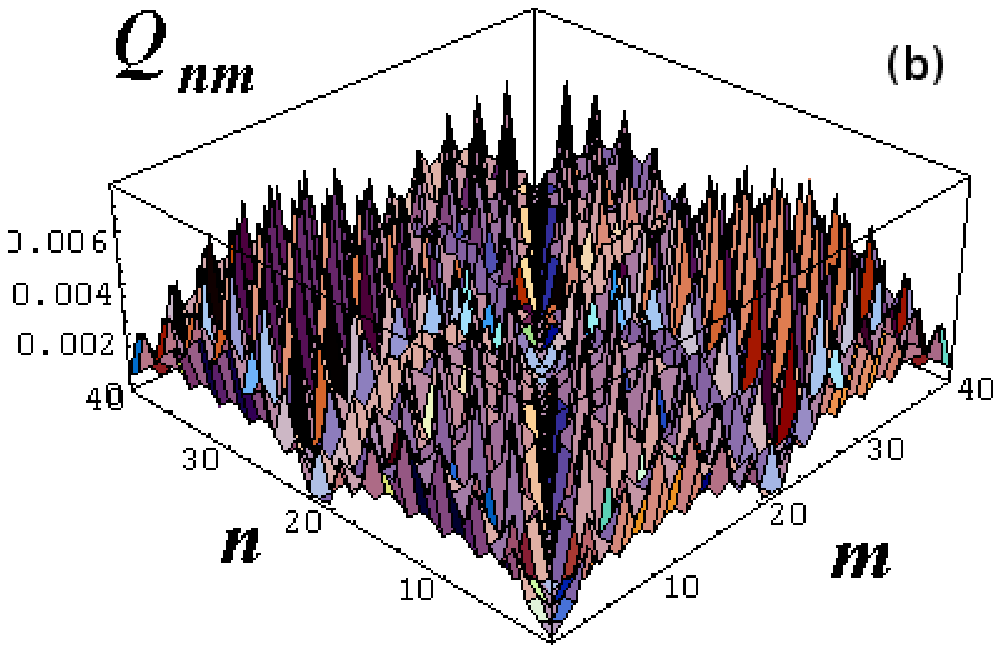,
    scale=0.8
   ,angle=0
}
  \caption{The stationary  pairwise discord $Q_{nm}$ among the virtual
particles in   the alternating  spin-1/2 chain with $N=41$, $\delta=0.1$; (a)
the distribution of  the discord for the initially excited node    $j=14$; (b) 
  the distribution of  ${\mbox{abs}}(Q_{nm}|_{j=14} -Q_{nm}|_{j=13} )$}
  \label{Fig:disc_hom01} 
\end{figure*}

As was mentioned in Sec.\ref{Section:alt_ch},
the diagonalization of the alternating XY Hamiltonian $H$ describing the nearest
neighbor interaction approximation    may be performed analytically 
\cite{FR,KF}. Remember also that, for even initially excited nodes $j$, the
discord/entanglement distribution coincides with that obtained for the
homogeneous spin-1/2 chain. 
 The basic novelty of the alternating chain is that related with the small
dimerization parameter $\delta$. We  verify the  conclusion of
Sec.\ref{Section:alt_ch}    that the discord distributions corresponding to $j=2
i$ and $2i-1$ are very similar, $i=1,2,\dots,20$. As an example, in
Fig.\ref{Fig:disc_hom01}a,
 we represent the discord distribution corresponding to $j=14$ for the small
dimerization parameter  $\delta=0.1$. To confirm that this distribution is very
similar to  the  distribution found for  $j=13$, we turn to Fig.
\ref{Fig:disc_hom01}b, where the   distribution of the absolute values of the
differences between both discords,
 ${\mbox{abs}}(Q_{nm}|_{j=14}-Q_{nm}|_{j=13})$,   is depicted. In both cases, the
strongly correlated   nodes are  $3 i-1$, $3i-2$, $i=1,2,\dots$. 
 Again, taking into account the DDIs among all nodes we only deform the
stationary pairwise discord distribution.

\subsubsection{3-alternating chain}

Consider the
3-alternating chain $d_{3i+1} =d_1\equiv 1$, $d_{3i+2}=d_2=1/2$ and
$d_{3i}=d_3=1/4$, $i=0,1,2,\dots,13$ \cite{K_F}.
We show only the discord distributions  corresponding to such excited nodes $j$
that reveal some new features  of the spin cluster with respect to the clusters
in the homogeneous chain.

\begin{figure*}

\epsfig{file=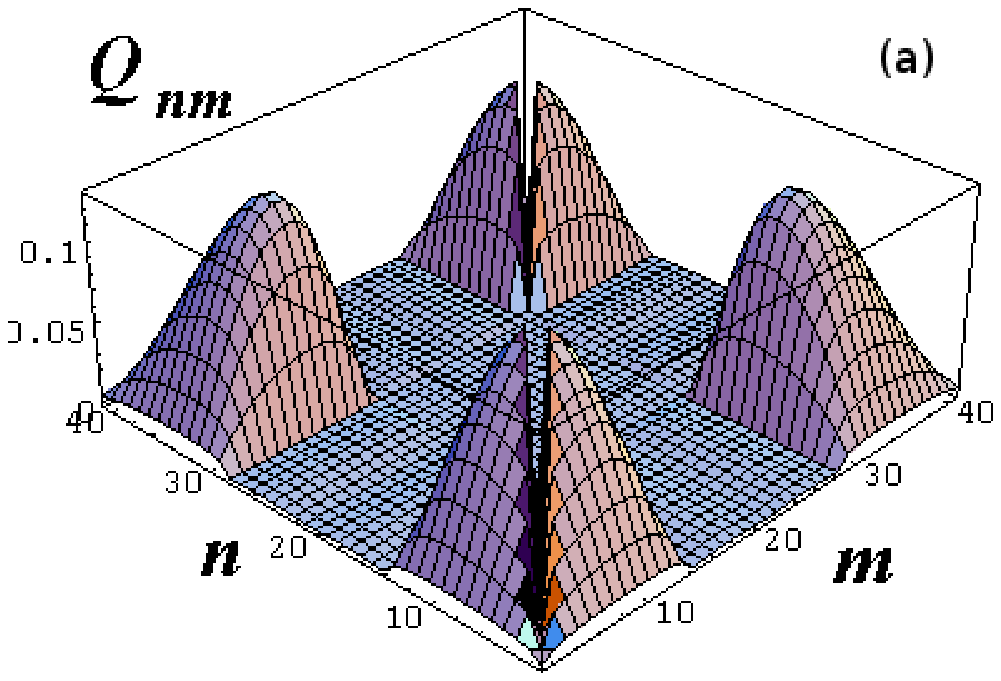,
    scale=0.8
   ,angle=0
}
\epsfig{file=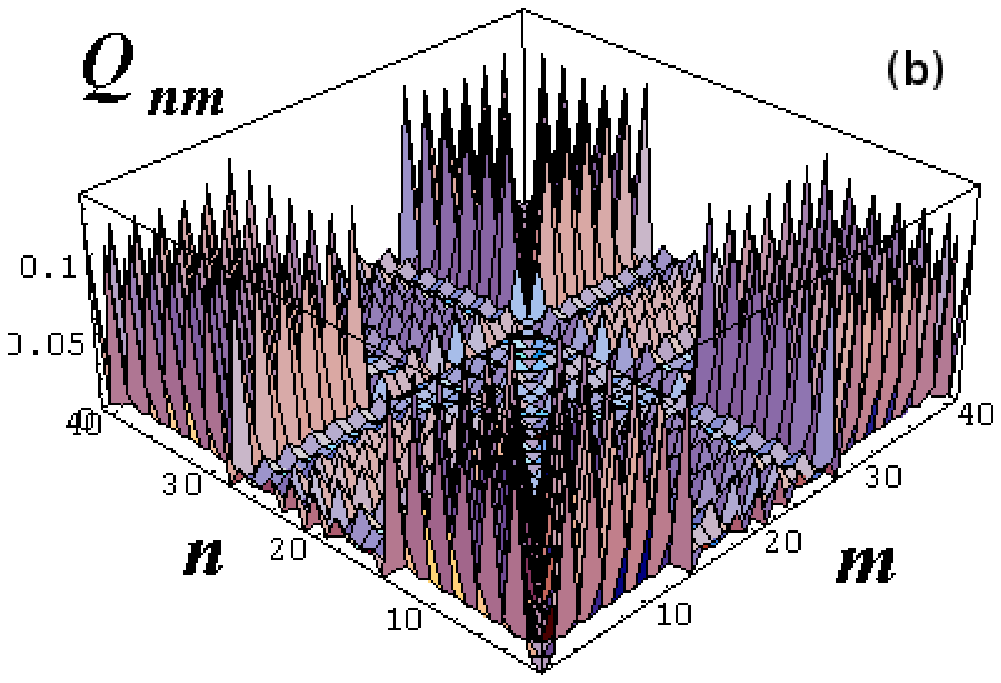,
    scale=0.8
   ,angle=0
}
\\\epsfig{file=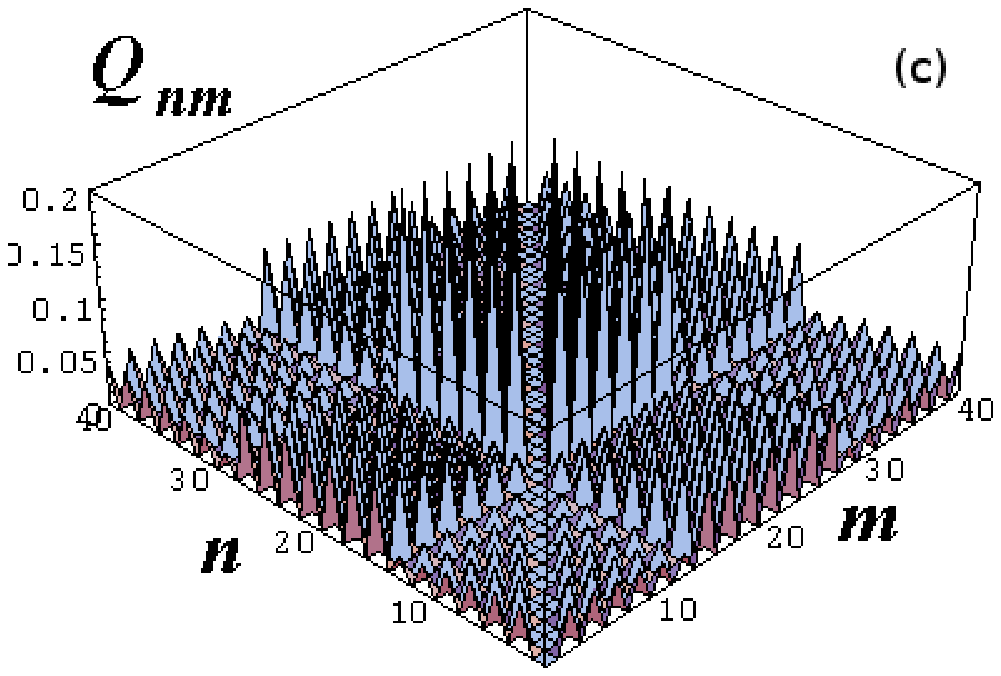,
    scale=0.8
   ,angle=0
}
\epsfig{file=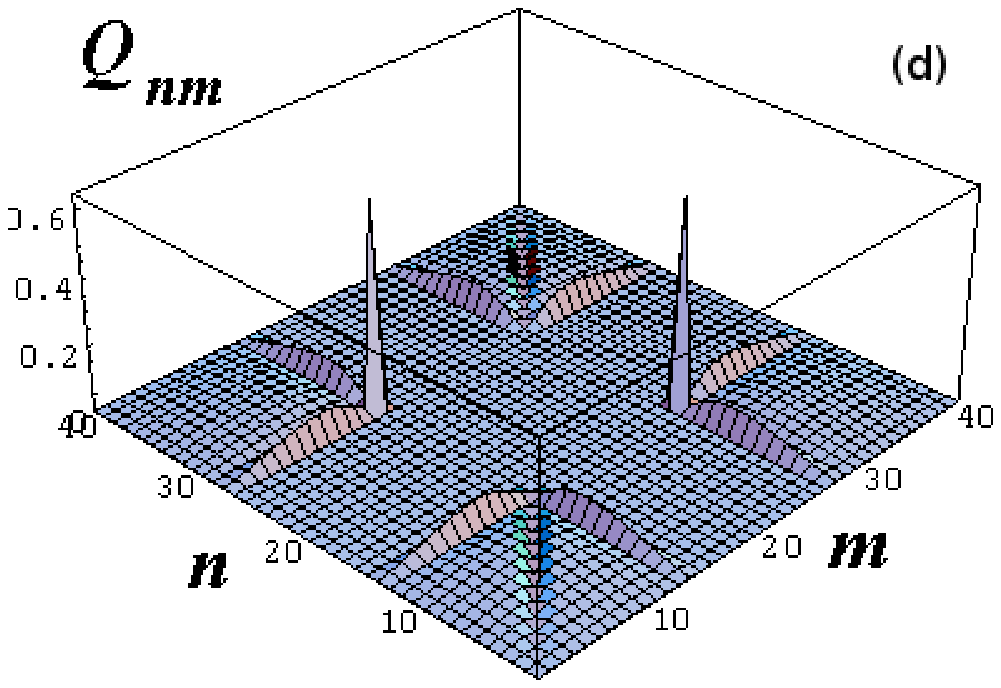,
    scale=0.8
   ,angle=0
}
  \caption{  The distribution of the stationary  pairwise discord $Q_{nm}$ among
the virtual particles in   the 3-alternating  spin-1/2 chain with $N=41$
   at the approximation of the nearest neighbor interactions. The initially
excited spins  are  following: (a) $j=2$, nodes  $k=14-28$ are excluded from the
cluster of correlated virtual particles, (b) $j=20$, 
  the cluster involves nodes  $k=1,3,5,7,9,11,13,15,29,31,33,35,37,39,41$;
   (c) $j=21$,  the cluster involves nodes $k=15,17,19,21,23,25,27$; (d) $j=40$,
nodes $14$ and $28$ are strongly correlated between themselves}
  \label{Fig:disc_alt3} 
\end{figure*}

1. $j=2$,   nodes  $n=14-28$ are excluded from the cluster of correlated
virtual particles (i.e. from the cluster with non-vanishing pairwise discord),
see Fig.\ref{Fig:disc_alt3}a

2. $j=20$,   strongly correlated  nodes (i.e. the  nodes with the pairwise
discord significantly larger then the pairwise discord between other nodes) are
$n=1,3,5,7,9,11,13,15,29,31,33,35,37,39,41$, see
Fig.\ref{Fig:disc_alt3}b

3. $j=21$,   strongly correlated  nodes are $n=15,17,19,21,23,25,27$, see
Fig.\ref{Fig:disc_alt3}c

4. $j=40$, strongly correlated  nodes are only two nodes $n=14$ and $n=28$, see
 Fig.\ref{Fig:disc_alt3}d
 
Thus, adding one more parameter (the coupling constant $d_3$) allows us to
create  additional types of the correlated clusters.

\subsubsection{Symmetric chain with 
$ \displaystyle d_i=\sqrt{\frac{i (N-i)} {{N-1}}} $, $1\le i \le 20$
\cite{CDEL}}
 Considering other variants of alternating chains we may achieve a large variety
of different clusters of virtual particles. We represent one more example of
the spin chain introduced  in \cite{CDEL} for the purpose of realization of  
 the perfect state transfer along the long spin-1/2 chain governed by the XY
Hamiltonian at the nearest neighbor interaction approximation. It is remarkable
that the dimensionless time interval needed for the perfect  excited state
transfer between end nodes does not depend on the length of the chain and equals
 $\pi$. 

 The most interesting pairwise discord distributions  are following.

\begin{figure*}
\epsfig{file=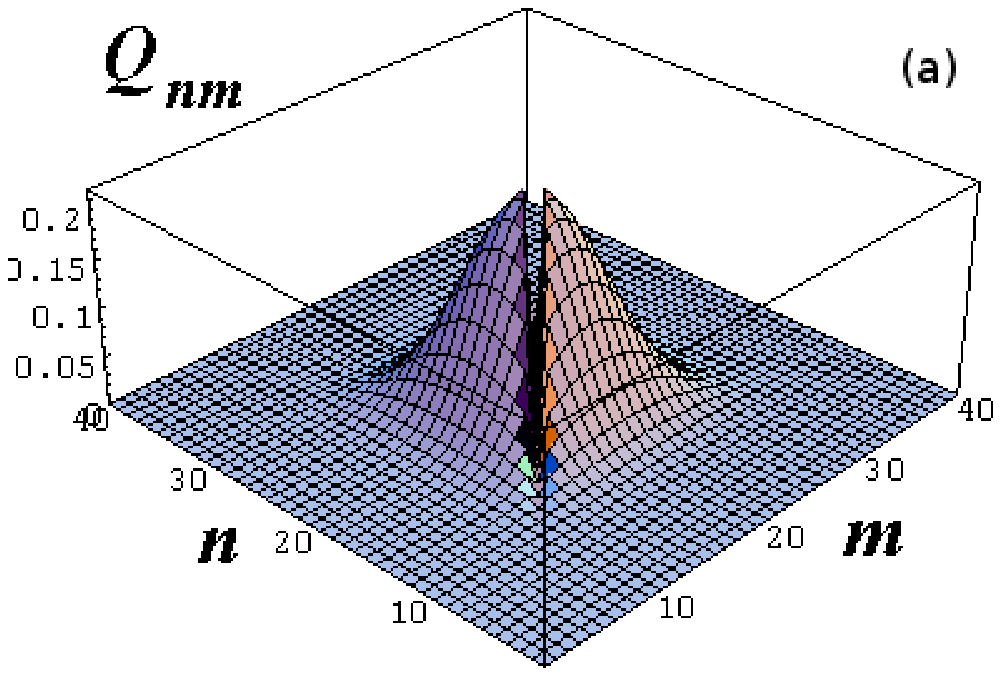,
    scale=0.8
   ,angle=0
}
\epsfig{file=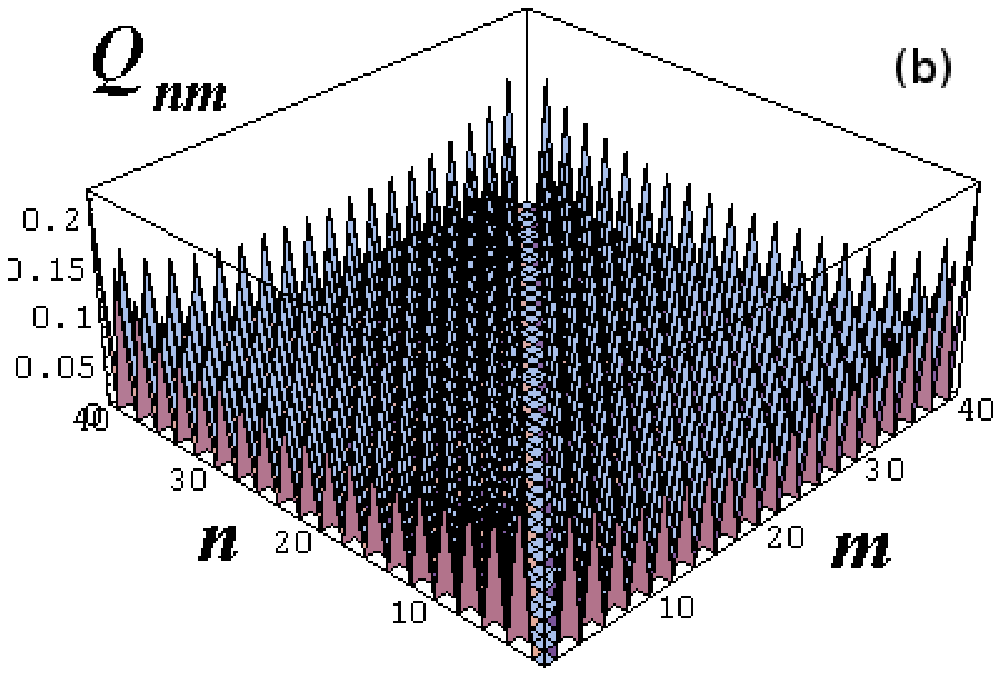,
    scale=0.8
   ,angle=0
}
  \caption{ The distribution of the stationary  pairwise discord $Q_{nm}$ among
the virtual particles in   
  the symmetrical  spin-1/2 chain, $d_i= \sqrt{\frac{i (N-i)}{N-1}} $, $1\le i
\le 20$,  with $N=41$ at the nearest neighbor interaction approximation; the
initially  excited nodes are following: (a) $j=1$, the distribution is bell
shaped with smaller cluster of the correlated nodes than in the homogeneous
chain with $j=1$, Fig.\ref{Fig:disc_hom}; (b) $j=21$, the family  of odd nodes
forms the cluster of correlated nodes}
  \label{Fig:disc_symE} 
\end{figure*}

1. $j=1$, the cluster of nodes $n=11-29$ is formed, Fig.\ref{Fig:disc_symE}a,
which is similar (but smaller) to the cluster in Figs.\ref{Fig:disc_hom}a and
\ref{Fig:disc_alt3}a

2  $j=21$, the cluster of odd nodes is formed, Fig.\ref{Fig:disc_symE}b,
but, unlike the homogeneous chain with the initially excited spin $j=21$, the
pairwise discord is not the same for all pairs.

\subsection{Spin-1/2 chains with a single initially  polarized node}
\label{Section:chain}
As  mentioned in Sec.\ref{Section:pol}, this initial state is more realistic and
may be created at  high temperatures. However, one has to remember the overall
effect of the temperature on the value of the discord and entanglement. It is
well known that both vanish with the increase in the temperature, i.e. the
quantum correlations are significant only at low temperatures. In our
calculations we take the dimensionless inverse temperature $\beta=10$.

Notice that  the distribution of the pairwise discord among the virtual
particles is very similar to
that obtain for the single initially excited  spin in
Sec.\ref{Section:excited_num}. 
This fact simplifies study of the discord for this more practically realizable
initial state. Nevertheless, we underline basic differences of the discord
distribution in this case to show that initially polarized  state can be 
preferable in some situations. In addition, unlike the initial state with the
single excited spin, the entanglement is identical to zero for long chains  in
this case, which was  proven in \cite{FZ_2012} for the homogeneous chains with
$N>4$.  
This fact is in favor of the discord as a measure of quantum correlations
revealing more quantum properties than the entanglement.

\subsubsection{Homogeneous chain}

\begin{figure*}
   \epsfig{file=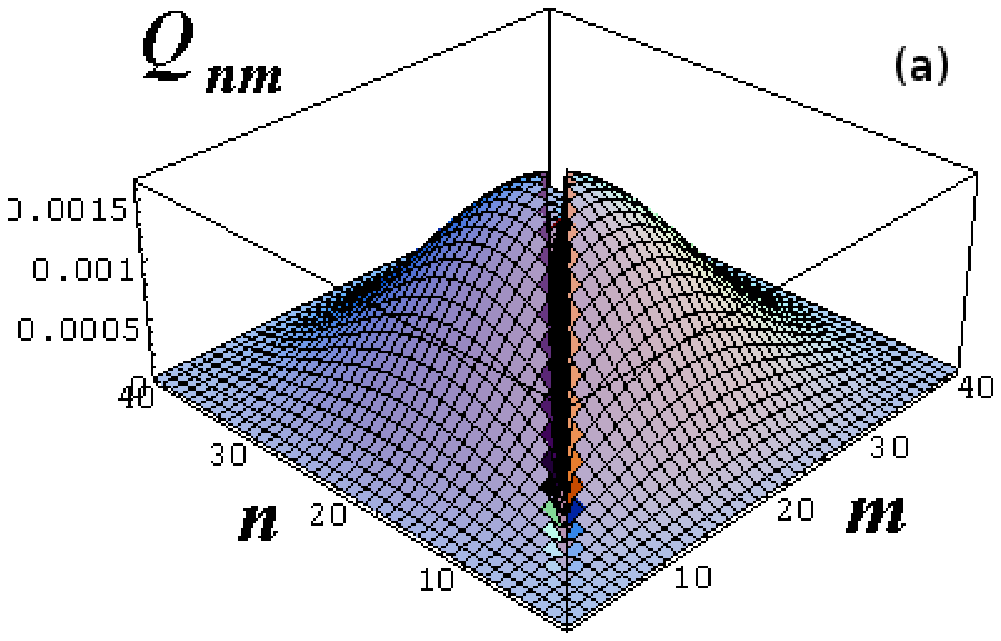,
    scale=0.8
   ,angle=0
}
\epsfig{file=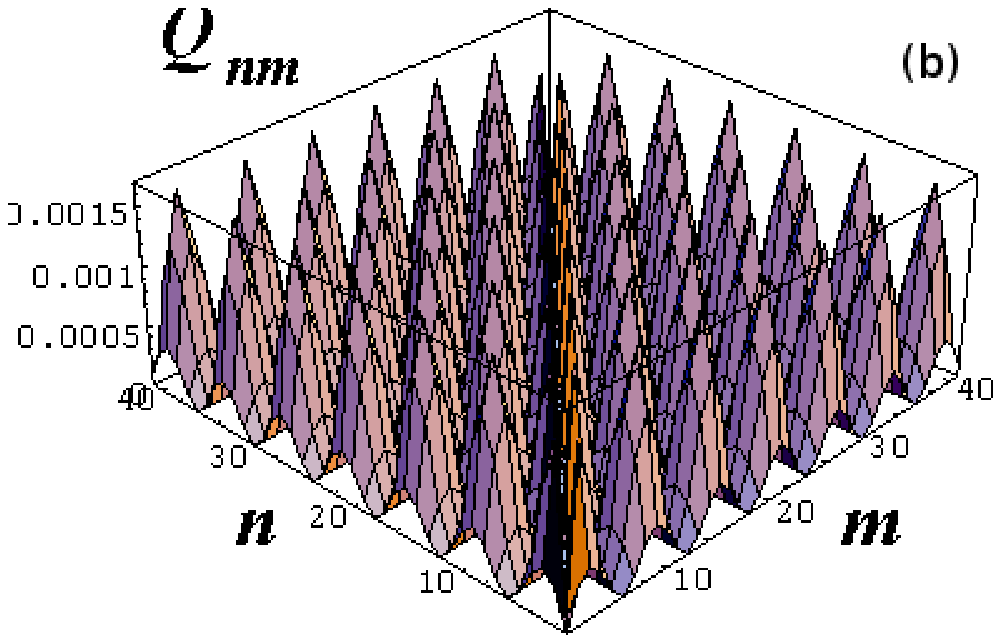,
    scale=0.8
   ,angle=0
}
  \caption{ The distribution of the stationary  pairwise discord $Q_{nm}$ among
the virtual particles in   
  the homogeneous spin chain-1/2 with $N=41$ at the nearest neighbor interaction
approximation. The initially polarized nodes are (a) $j=1$, (b) $j=7$. In both
cases the shapes of destructions are similar to those obtained  for the
homogeneous chain, see Fig.\ref{Fig:disc_hom}}
  \label{Fig:disc_homECHO} 
\end{figure*}

First, we consider the homogeneous chain (Fig.\ref{Fig:disc_homECHO}) and
compare the discord distribution with that obtained in
Sec.\ref{Section:1excited}, Fig.\ref{Fig:disc_hom}.

Obviously, 
the shapes of the discord distributions are similar in both cases
with the following quantitative differences.

1. The initially polarized node $j=1$, Fig.\ref{Fig:disc_homECHO}a. All nodes
are correlated, and the minimum pairwise  discords correspond to pairs involving
the end nodes. Comparing Figs.\ref{Fig:disc_homECHO}a and \ref{Fig:disc_hom}a 
we conclude that     the discord is steeper in the case of initially  polarized
node, which means that the edge nodes are less correlated with the center nodes 
in this case. Consequently, the center nodes are better correlated with each
other and  the edge  nodes are less sensitive to the perturbations of the center
nodes. These perturbations must be large enough to effect the edge nodes.  

2. If the initially polarized node is $j=7$, we have (see
Fig.\ref{Fig:disc_homECHO}b)
three nonzero values for  the elements $\rho_{nn}$, (i.e. $\rho_i$, $i=1,2,3$) 
given  by eq.(\ref{rhoexj7}) and, consequently, three different values $J_i$,
$i=1,2,3$, for $J_{nn}$:
\begin{eqnarray}\label{Jj7}
&&
J_i=\frac{\tanh \,5}{2} \rho_i,\;\;i=1,2,3.
\end{eqnarray}
Therewith
\begin{eqnarray}\label{Qpol}
&&
Q^{pol}_1=Q^{pol}_{15}\approx 0.00010,\;\;Q^{pol}_2=Q^{pol}_{24}\approx
0.00092,\;\;
Q^{pol}_3=Q^{pol}_{39}\approx 0.00164,\\\nonumber
&&
Q^{pol}_4=Q^{pol}_{12}\approx 0.00031,\;\;Q^{pol}_5=Q^{pol}_{13}\approx
0.00041,\;\;
Q^{pol}_6=Q^{pol}_{23}\approx 0.00123
\end{eqnarray}
All nodes are correlated accept for the nodes $n=6i$, $i=1,\dots,6$ because all
the pairwise discords involving these nodes  are zeros.

Notice that 
the  spread of the discord (i.e. the ratio of the  difference between the
maximum  and minimal nonzero discords   to the maximal discord) is larger in the
case of a single polarized node, which follows from the comparison of
Figs.\ref{Fig:disc_hom}b and \ref{Fig:disc_homECHO}b. This means, in particular,
that the nodes from the set  $\{6 i +1, 6 i +5,\;\; i=0,1,\dots,6\}$ are  less
sensitive to the perturbations of other virtual particles in the case of the
initially polarized 7th node.

3. If the initially polarized node is $j=14$, we have the only nonzero value
$\rho_{nn}$ (i.e. $\rho_1$) given by eq.(\ref{rhoexj14}) and
\begin{eqnarray}
&&J_1=\frac{\tanh \,5}{2} \rho_1=\frac{\tanh \,5}{56} ,\\\nonumber
&&
Q^{pol}_1=0.00092,\;\;n=3i-1, 3i-2, \;\;i=1,\dots,13
.
\end{eqnarray}
The cluster of correlated nodes is formed by the nodes 
$\{3 i-1,3i-2; \; i=1,2,\dots,13\}$.

4. If $j=21$, then the only nonzero value of $\rho_{nn}$ (i.e. $\rho_1$) is
given by eq.(\ref{rhoexj21}) and
\begin{eqnarray}
&&
J_1=\frac{\tanh \,5}{42},\\\nonumber
&&
Q^{pol}_1=0.00164\;\;n=2i -1,\;\;i=0,1,\dots,21 .
\end{eqnarray}
Thus, the cluster of correlated nodes is represented by  the family  of odd
nodes.

In general,  the absolute value of the discord is significantly less in the case
of a single initially polarized node, as follows  from the comparison of the
both graphs in Figs.\ref{Fig:disc_hom} and \ref{Fig:disc_homECHO}.

\subsubsection{Alternating chain}
\begin{figure*}

\epsfig{file=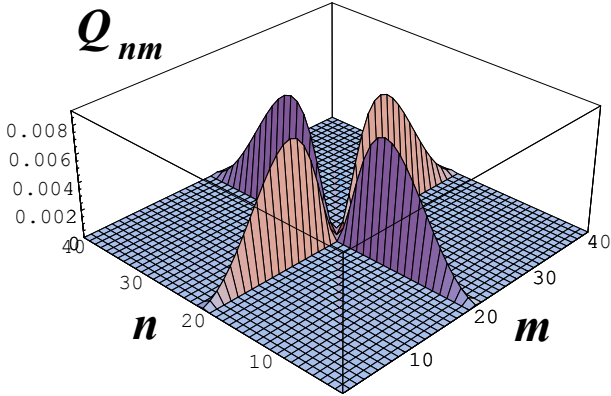,
    scale=0.8
   ,angle=0
}
  \caption{The distribution of the stationary  pairwise discord $Q_{nm}$
  in the alternating spin-1/2 chain with $N=41$, $\delta=1/2$ and the initially
polarized  node  $j=41$ at the  nearest neighbor interaction approximation. The 
  21th node correlates with all others.
 }
  \label{Fig:disc_altECHO} 
\end{figure*}

Similar to Sec.\ref{Section:alt}, the discord does not depend on the
dimerization parameter $\delta$ if the initially polarized node 
$j$ is even. 
Comparing  this discord with the discord for the alternating chain with
initially excited node 
we conclude  that the same remarks as for the homogeneous chain are valid in
this case. Namely, the discord is steeper if $j=1$, the spread of the discord is
larger and the value of discord is smaller. A novelty is that at $j=41$, unlike
the discord in the  chain with the initially  excited spin $j=41$,
 the node $n=21$ is correlated with all other nodes, while other  correlations
are negligible, as demonstrated in Fig.\ref{Fig:disc_altECHO} for the chain with
$\delta=1/2$.

\section{Conclusions}
\label{Section:conclusions}
We study such representations of the density matrix associated with a quantum
system of spin-1/2 particles which reveal the stationary distributed pairwise
discord. This system is a system of virtual particles associated with the
eigenstates of the Hamiltonian. In particular, if the nearest neighbor
interaction approximation is used, this system of virtual particles is the
system of $\beta$-fermions \cite{FZ_2012}. The systems with the stationary
discord are convenient for the realization of the quantum operations. In
addition, it is much simpler to prepare the desirable distribution of the
discord, since it does not evolves.  Using different coupling constants,
different initial states and different Hamiltonians governing the spin dynamics
we may handle the size of the cluster of coherent particles. 

Emphasize, that the above virtual particles are not localized in the physical
space, which 
creates a problem of "interaction" with these particles using the classical
tools. However, this problem disappears in so-called "inner" parts of quantum
algorithms where such interaction is absent. We assume that systems with
stationary discord are most suitable namely for 
 these algorithms.

Examples of homogeneous and non-homogeneous spin-1/2  chains  (alternating,
3-alternating and completely inhomogeneous chain of ref. \cite{CDEL}) are
considered with two types of the initial conditions: the single initially
excited and single initially polarized node. The peculiarity of the initial state
with the single initially excited node is that the both discord and entanglement are
non zero in the above system of virtual particles in this case.  
We found (both analytically and numerically) that  the stationary 
discord/entanglement distribution is defined by the position of the initially
excited/polarized 
node. It is interesting that the shapes of discord and  entanglement
distributions are essentially the same in the case of initially excited node. In
addition, this shapes remain essentially the same for the discord distribution
in the case of initially polarized node. 
Set of  peculiar subsystem of  correlated virtual particles have been found.
It is important that the subsystems of large numbers of virtual particles with
equal pairwise discord are among them. Such subsystems might be  proper
candidates for  the quantum registers.

It is shown that the remote DDIs  only slightly deform the distribution of the
discord in a quantum system, unlike the evolution of the pairwise  quantum
correlations in the system of physical spins  and the state transfer process 
along the 
spin chains, which significantly depend on the interactions among remote nodes.

This work is supported by the Program of the Presidium of RAS No.8 ''Development
of methods of obtaining chemical compounds and creation of new materials'' and
by the Russian Foundation for Basic Research,   grant 
No.13-03-00017. 

\section{Appendix: Minimization in eq.(\ref{CB2}) }
\label{Section:app}

 Let us show that the minimum in eq.(\ref{CB2}) corresponds to $\eta=0$, similar
to ref.\cite{FZ_2012}. Eqs.(\ref{p}) and (\ref{theta}) 
 at $\eta=0$ yield
 \begin{eqnarray}
 &&
 p_i(0) \equiv p_i|_{\eta=0}=\frac{1}{2},\\
 &&
 \theta_i(0)\equiv \theta_i|_{\eta=0}=2 \sqrt{\rho_{nn}\rho_{mm}
+\frac{1}{4}(1-2\rho_{mm})^2},\;\;i=0,1.
 \end{eqnarray}
 Consequently, using the definition of $S_i$ given by eq.(\ref{S}), we conclude
that  $S_1|_{\eta=0}=S_0|_{\eta=0} \equiv S(\theta_0(0))$ and
 \begin{eqnarray}
 (p_0S_0 +p_1 S_1)|_{\eta=0} = 2 p_0(0)  S(\theta_0(0))=  S(\theta_0(0)) =
 S\left( 2 \sqrt{\rho_{nn}\rho_{mm} +\frac{1}{4}(1-2\rho_{mm})^2} \right)
 \end{eqnarray}
 
Similarly, Eqs.(\ref{p}) and (\ref{theta})  at $\eta=1$ yield
 \begin{eqnarray}
 &&
 p_0(1)=1-\rho_{nn},\;\; p_1(1)=\rho_{nn},\\
&&
\theta_0(1)= \frac{|1-2\rho_{mm}-\rho_{nn}|}{1-\rho_{nn}}
 \\
 &&\theta_1(1)=1
 \end{eqnarray}
Again, using the definition of $S_i$ given by eq.(\ref{S}) we have
$S_1|_{\eta=1}=0$ and
 we can write
 \begin{eqnarray}
 (p_0S_0 +p_1 S_1)|_{\eta=1} =  p_0(1) S(\theta_0(1)) = (1-\rho_{nn})
S\left(\frac{|1-2\rho_{mm}-\rho_{nn}|}{1-\rho_{nn}}\right).
 \end{eqnarray}
 Thus we have to find the minimum of two quantities:
 \begin{eqnarray}
 \min\left(
 S\left( 2 \sqrt{\rho_{nn}\rho_{mm} +\frac{1}{4}(1-2\rho_{mm})^2} \right),
 (1-\rho_{nn}) S\left(\frac{|1-2\rho_{mm}-\rho_{nn}|}{1-\rho_{nn}}\right)
 \right)
 \end{eqnarray}
 Representing the ratio of these two quantities as a two-dimensional surface in
the space of the parameters $\rho_{nn}$ and $\rho_{mm}$ ($\rho_{nn},\rho_{mm}\le
1$, $\rho_{nn}+\rho_{mm} \le 1$) we conclude that the first of them
(corresponding to $\eta=0$) is always less than the second one.  Consequently 
the minimum in eq.(\ref{CB2}) is always at $\eta=0$.

\end{document}